\documentclass{PoS}

\newcommand{\Fermi}{\textit{Fermi}}
\newcommand{\FermiLAT}{\Fermi-LAT}

\newcommand{\hess}{\textsc{H.E.S.S.}}
\newcommand{\gr}{$\gamma$-ray}
\newcommand{\grs}{$\gamma$-rays}
\newcommand{\dgr}{\ensuremath{^\circ}}

\title{Intrinsic spectra  of H.E.S.S. blazars : what would we see without EBL absorption}

\ShortTitle{H.E.S.S. Intrinisic Spectra}

\author{C. Romoli\\
                Dublin Institute for Advanced Studies, 31 Fitzwilliam Place, Dublin 2, Ireland}
\author{\speaker{D. A. Sanchez}\\
                LAPP,
                Univ. de Savoie, CNRS/IN2P3, Annecy-le-Vieux F-74941, France \\
                E-mail: \email{david.sanchez@lapp.in2p3.fr}}
\author{M. Lorentz\\
                DSM/Irfu, CEA Saclay, F-91191 Gif-Sur-Yvette Cedex, France}
\author{P. Brun \\
                DSM/Irfu, CEA Saclay, F-91191 Gif-Sur-Yvette Cedex, France}
\author{on behalf of the H.E.S.S. collaboration}

\abstract{The vast majority of extragalactic sources detected in the very high
energy (E> 100 GeV) domaine are active galactic nuclei (AGN) located at
cosmological distances. During their travel towards Earth, the emitted
gamma-rays suffer from absorption by the extragalactic background light
(EBL). The density of the EBL is not very well constrained by direct or
indirect measurement which leads to uncertainties on the intrinsic
spectrum of the sources. High-quality AGN spectra obtained with the
High Energy Stereoscopic System (H.E.S.S.) have been used to perform a
model-independent measurement of the EBL spectral energy distribution.
While the precision of this measurement  remains limited, it reflects
the sensitivity of H.E.S.S. alone to the EBL and offers the possibility
to access the intrinsic spectra of AGNs in a consistent manner, taking
into account the derived uncertainties on the EBL spectral energy
distribution.

In this contribution, we study the intrinsic spectra as measured by
H.E.S.S. and by the Fermi Large Area Telescope (LAT) of several blazars
of the H.E.S.S. sky. This provides the opportunity to have new
insight into the emission processes at play in the jets of AGN. The data
presented consist of monitoring data of quiescent state of blazars and
also bright blazar flares, such as PKS~2155-304, Mrk~421, recorded
by H.E.S.S.}

\FullConference{35th International Cosmic Ray Conference – ICRC217-\\
		10-20 July, 2017\\
		Bexco, Busan, Korea}

\begin{document}

\section{Introduction}

While traveling cosmological distances, the VHE \grs\ coming from blazars interact with lower
energy photons from the extragalactic background light (EBL), resulting in an
energy- and redshift-dependent absorption. The density of EBL photons is subject to uncertainties, direct measurements
suffering from the contamination of foreground sources \cite{2013APh....43..112D} (e.g. zodiacal light),
and galaxy numbers counts being only considered as lower limits
\cite{SummaryObs2}. As a consequence, the intrinsic VHE spectra of distant
sources is still unknown.

The EBL absorption leaves a clear imprint on the intrinsic source spectrum,
forming the basis of the first detection of the EBL  with the \hess\ telescopes
using bright or distant blazars \cite{2013A&A...550A...4H}. This measurement,
recently updated by \cite{ebl}, allowed to probe the intrinsic spectral
properties, in the VHE range, of the sources used to derive the EBL measurement
in a coherent manner.

\section{Analysis}

In this study, the same data set as in \cite{ebl} is used and the local EBL
density considered for the correction of the spectra is also derived in
\cite{ebl}. When fitting together \FermiLAT\ and \hess data, only
contemporaneous \Fermi\ data were considered with the \hess  data set. These
were split into flux bins in \cite{ebl}, and have been merged. Faint \Fermi\
sources have their spectra derived with 8 years of data. Time ranges of the
\FermiLAT\ analysis are given in Table \ref{table:list}.

\subsection{H.E.S.S data sets}

The High Energy Stereoscopic System (\hess) is located in the Khomas
Highlands, Namibia (23\dgr16'18'' S, 16\dgr30'01'' E), at an altitude
of 1800 m above sea level. In its first phase, \hess\ was an array of
four identical imaging atmospheric Cherenkov telescopes. In 2012, a
fifth telescope has been added in the center of the array. This study
only considers data taken during the first phase.

A large part of the \hess\ observation time is devoted either to the detection of
new blazars, or the monitoring of well known VHE sources or to
target of opportunity pointings. The analyzed sources and the corresponding data
sets have been presented in \cite{ebl} together with the analysis procedure. In
total, 9 sources are included in this study. Name, coordinates and redshift of
the sources are given in Table \ref{table:list}.

Mrk~421 has been observed by H.E.S.S. in 2004 (3 data sets) and during
a flaring episode in 2010 \cite{2010tsra.confE.197T} (2 data sets). All
the observations taken on PKS~2005-489 have been divided into two flux bins to
ensure sufficient statistics. PKS~2155-304 is extensively observed by H.E.S.S. since
the beginning of the experiment which recorded an exceptional flare in 2006
\cite{2007ApJ...664L..71A}. This event has been divided in 7 flux bins and the
data taken in 2008 during a multi-wavelengh campaign \cite{2009ApJ...696L.150A}
makes the last data set for this source.

\begin{table*}[p]
\caption{List of source (Name, RA, Dec and redshift) used in this study. The
last column gives the time range of the \FermiLAT\ analysis in MET (Mission
Elapsed Time), see section \ref{fermi}. } 
\label{table:list} 
\centering %
\begin{tabular}{|l|l|l|l|l|}\hline
   Name & RA & Dec & redshift & \Fermi\ time range \\\hline\hline
   1ES 0229+200 & 02 32 53.2 &+20 16 21 & 0.1396 & 239557418.0 - 491961604.0  \\
   1ES 0347-121 & 03 49 23.0 & -11 58 38 & 0.188& 239557418.0 - 491961604.0\\
   1ES 0414+009 & 04 16 52.9 & +01 05 20&0.287& 239557418.0 - 491961604.0\\
   1ES 1101-232 & 11 03 36.5 & -23 29 45 &0.186& 239557418.0 - 491961604.0\\
   Markarian 421 &11 04 19  & +38 11 41	 &	 0.031& 288057602.0 - 288403202.0 \\
   1ES 1312-423	 & 13 14 58.5 & -42 35 49 &0.105& 239557418.0 - 491961604.0\\
   PKS~2005-489 & 20 09 27.0 & -48 49 52 & 0.071& 239557418.0 - 491961604.0 \\
   PKS~2155-304 & 21 58 52.7 & -30 13 18 & 0.116& 241315201.0 - 242438401.0\\
   H 2356-309 & 23 59 09.4	& -30  37 22 & 0.165 & 239557418.0 - 491961604.0\\\hline

\end{tabular}
\end{table*}

Spectral points were extracted using a Bayesian unfolding technique
\cite{2007NIMPA.583..494A,2013PhRvD..88j2003A} allowing them to be obtained
independently of any spectral model, along with as the correlation matrix
between points. Points are then corrected by the EBL model as obtained in
\cite{ebl} and fitted taking into account the correlation matrix. The
uncertainty derived on the EBL photon density is added in quadrature to the
statistical errors of the \hess data.

\subsection{\FermiLAT\ analysis} \label{fermi}

The large area telescope (LAT) on-board the \Fermi\ satellite is a pair
conversion detector with a silicon strip tracker on the top of a
calorimeter \cite{2009ApJ...697.1071A}. A segmented anti-coincidence
shield allows the rejection of the charged particles. The bulk of LAT
observations are performed in an all-sky survey mode allowing to
observe all parts of the sky for about 30 minutes every 3 hours.

Data and software are publicly available from the Fermi Science Support
Center (FSSC). Each source has been analyzed in the same way using the
{\tt Enrico} Python package \cite{2013arXiv1307.4534S} adapted for {\tt
PASS 8} analysis. A region of interest (ROI) of 15\dgr\
radius, centered on the source was defined to extract the spectral
parameters. The {\tt PASS 8} data  (event class 128 and event
type 3) were used together with the corresponding response functions
{\tt P8R2\_SOURCE\_V6}. Note that, the time range of the analysis is
given in Table \ref{table:list}.  In addition, cut on the zenith angle
($<90^{\circ}$) was applied to remove the Earth albedo.

The sky model has been created including all the sources of the 3FGL
\cite{3FGL}, adding the Galactic  diffuse emission using the file {\tt
gll\_iem\_v06.fits} \cite{2016ApJS..223...26A} and the isotropic
background using {\tt iso\_P8R2\_SOURCE\_V6\_v06.txt}.

\section{Derivation of the spectral parameters }

In order to derive the spectral parameters of the sources, the  H.E.S.S. and \Fermi\
data points have been fitted in log-log space. In the H.E.S.S. energy range,
correlation between points (evaluated using the covariance matrix) is taken into
account. For this a simple $\chi^2$ fit was used and data have been fitted with
a power-law (LP), a log-parabola (LP) and a power-law with an exponential
cut-off (PLEC) model. The best model is chosen based on the values of the
$\chi^2$ with a cut at 5$\sigma$

\begin{figure} \centering
\includegraphics[height=.7\textwidth]{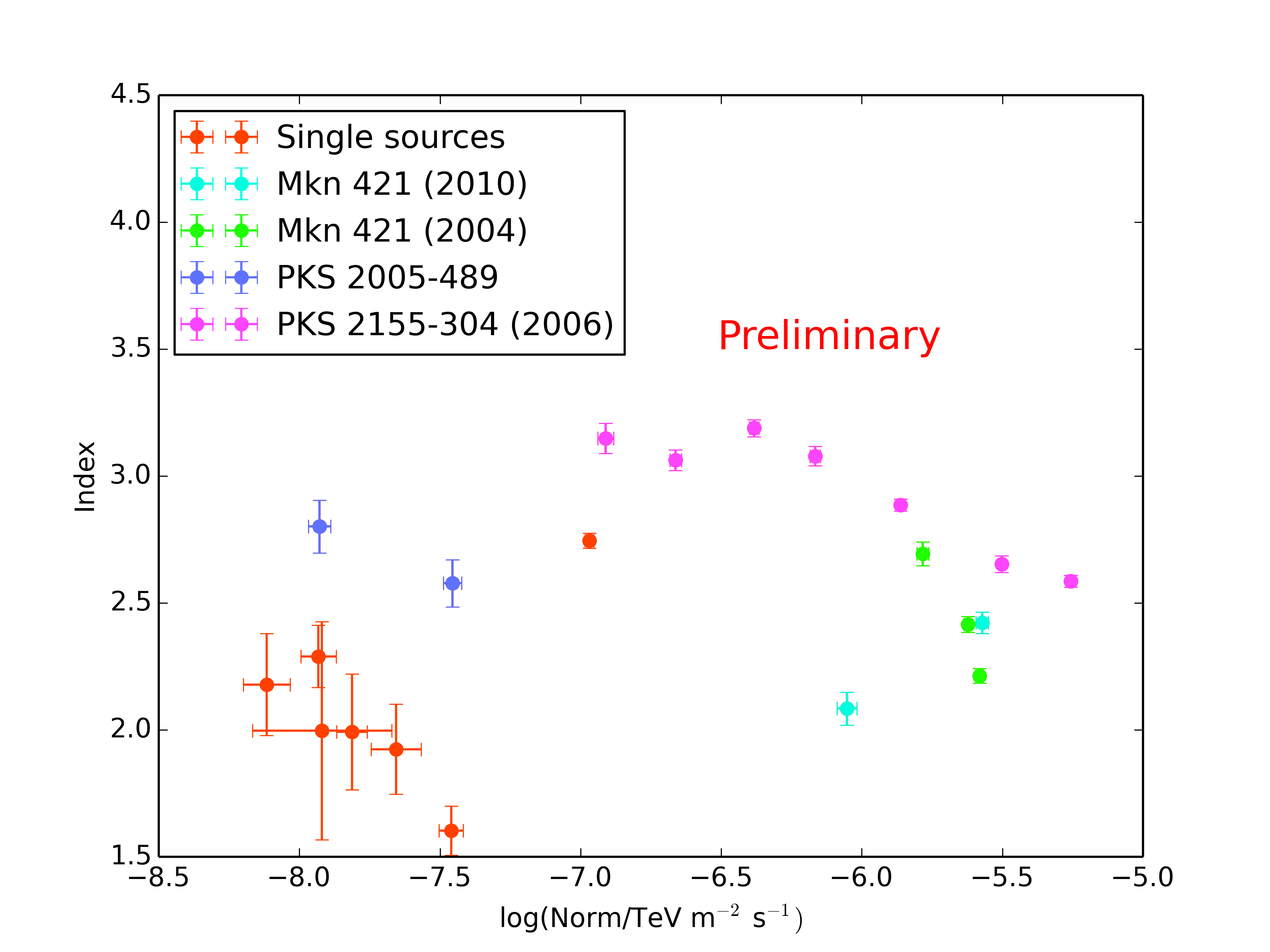}
\caption{Results of the fit of the unfolded \hess data points with a simple PL. The X
axis is the normalisation at 1 TeV and the Y axis is the spectral index $Gamma$.
Red points are the results for  1ES 0229+200, 1ES 0347-121, 1ES 0414+009, 1ES
1101-232, 1ES 1312-423, PKS~2155-304 (in 2008) and  H 2356-309. The bright state
of Mrk~421, PKS~2005-489 and PKS~2155-304 are shown separately. }
\label{HESSfit}
 \end{figure}

The first step is to only fit the \hess data. In this case, the vast majority of
the data set is well fitted with a PL. Figure \ref{HESSfit} show fitted spectral
index as a function of  the normalisation of the spectra at 1 TeV. 6 sources
exhibit a low flux together with a hard spectral index ($<2.5$ ) : 1ES
0229+200, 1ES 0347-121, 1ES 0414+009, 1ES 1101-232, 1ES 1312-423 and H 2356-309
(see section \ref{ul}).

During the 2006 flare, PKS~2155-304 exhibited spectral variability
\cite{2012A&A...539A.149H} which is confirmed here. Important spectral
variability is also found for Mkr 421 during the 2 flares of 2004 and 2010 recorded by \hess.
Interestingly the highest state in 2010 is compatible with the middle flux state
in 2014 but there is not enough data to draw any conclusion. PKS~2005-489 does
not seem to show spectral variability but here again, statistics are lacking for a
firm conclusion.

\subsection{Constraining the peak of the Spectral Energy Distribution }

Three sources have their \gr\ SED peak constrained by the \FermiLAT\ and \hess data :
Mrk~421, PKS~2005-489 (full data set) and PKS~2155-304 (2008 data set). The peak energy is in the GeV range
(Table \ref{table:max}). For Mrk~421, this result is obtained during a flaring
episode which is not the case of the 2 other sources. Nevertheless, the flux of
all the sources is higher than the flux of the other blazars of the sample (Fig   \ref{Emaxplot}).

\begin{figure}
    \resizebox{18pc}{!}{\includegraphics[height=.9\textheight]{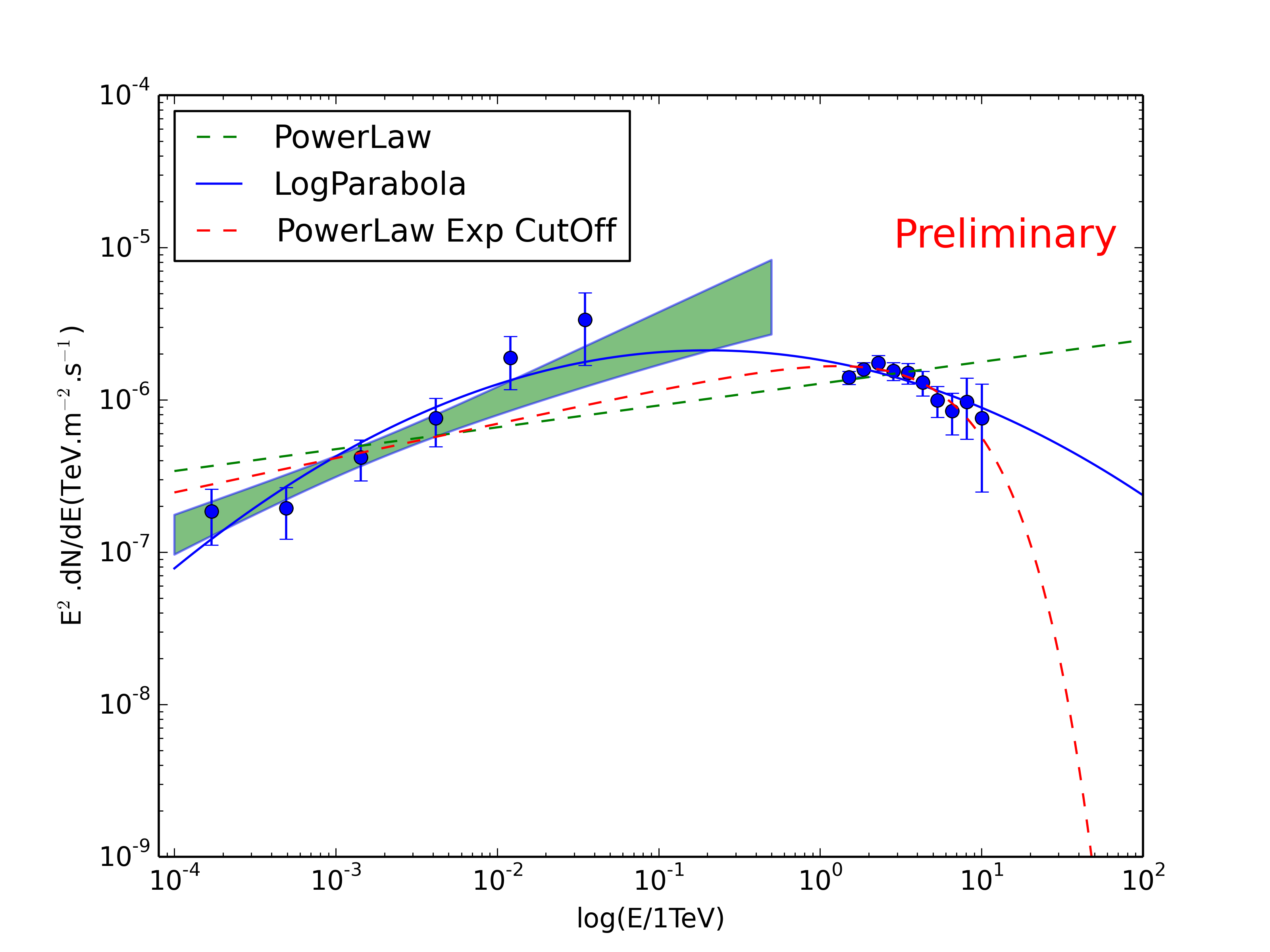}}
        \resizebox{18pc}{!}{\includegraphics[height=.9\textheight]{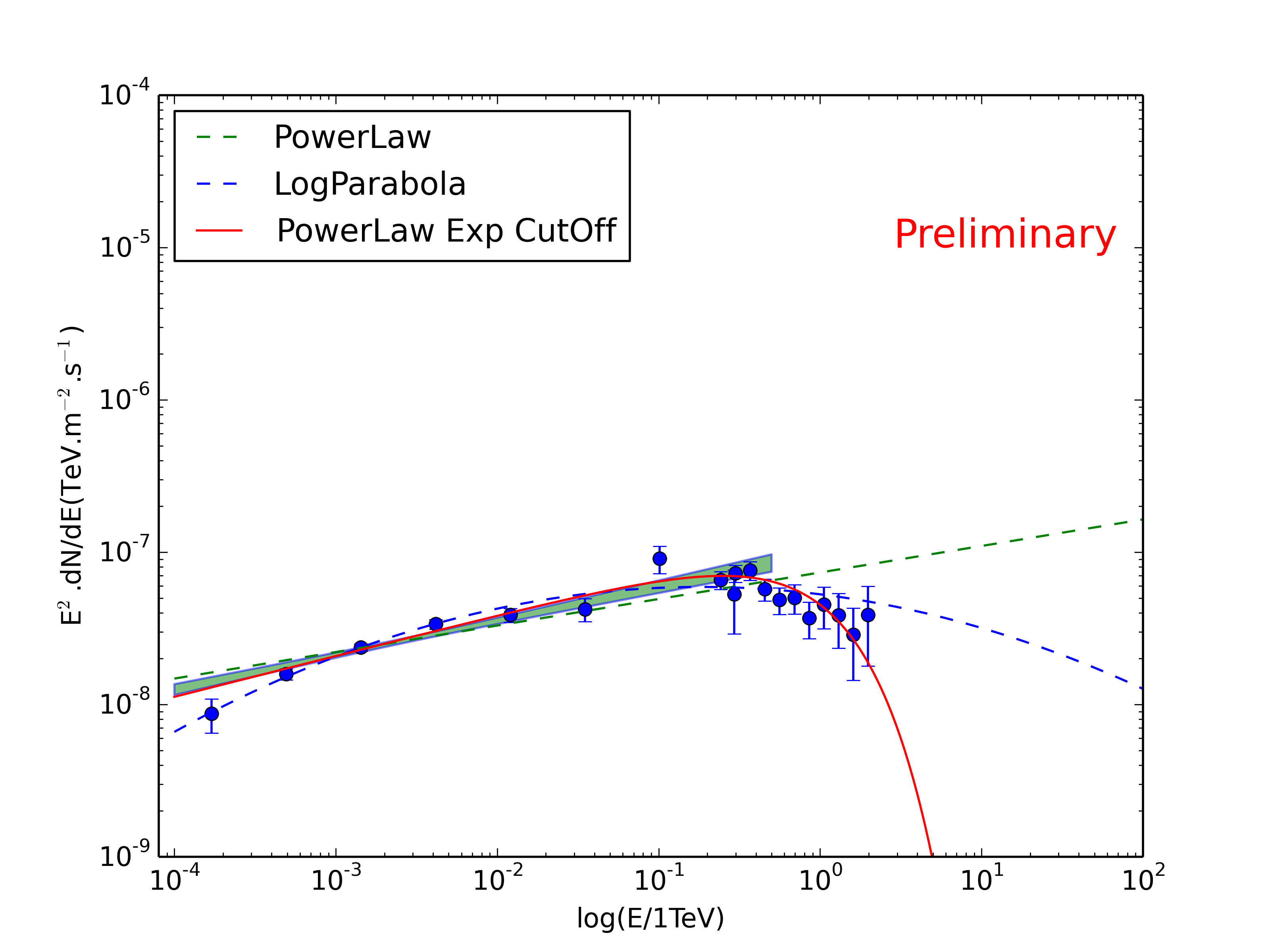}}
  \caption{SED of Mrk~421 (left) and PKS~2005-489 (right). Data points are the \FermiLAT\ and \hess data. The green butterfly is the 1$\sigma$ contour of the \FermiLAT\ analysis.
  Data were fitted with a PL, a LP and a PLEC. The best model is the one represented with a plain line (LP for Mrk~421 and PLEC for PKS~2005-489).}
\label{SEDMkr421}
\end{figure}

\begin{table*}[!hp]

\caption{Best fit model of the \grs\ data and peak position and flux value in
the $\nu f(\nu)$ representation of the sources which exhibit a significant turn
over in this representation.} 

\label{table:max} 
\centering %
\begin{tabular}{|l|l|l|l|}\hline
   Name & Model & $E_{\rm peak}$& $E^2 dN/dE_{\rm peak}$   \\
    &  &   [TeV] &  $10^{-7}$ [TeV.m$^{-2}$.s$^{-1}$]   \\\hline
   Mrk~421 & LP & 0.20$\pm$0.09 & 21.4$\pm$2.3  \\
   PKS~2005-489& PLEC & 0.25$\pm$0.02 & 0.70$\pm$0.10\\
   PKS 2155-304 & PLEC & 0.081$\pm$0.009 & 3.5$\pm$0.7 \\\hline
\end{tabular}
\end{table*}

\subsection{Bayesian limits on the peak position}\label{ul}

\begin{figure}
    \resizebox{18pc}{!}{\includegraphics[height=.9\textheight]{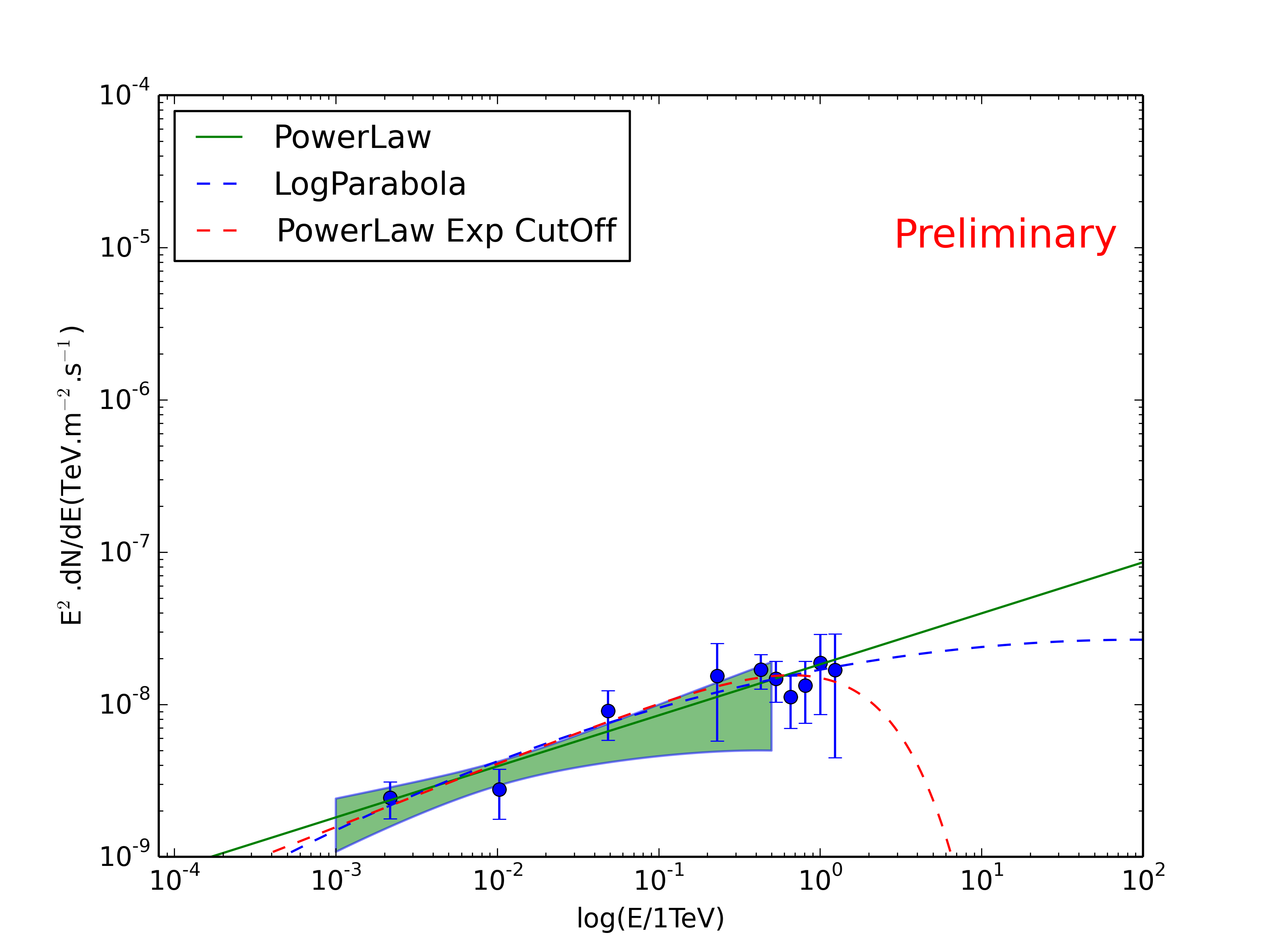}}
        \resizebox{18pc}{!}{\includegraphics[height=.9\textheight]{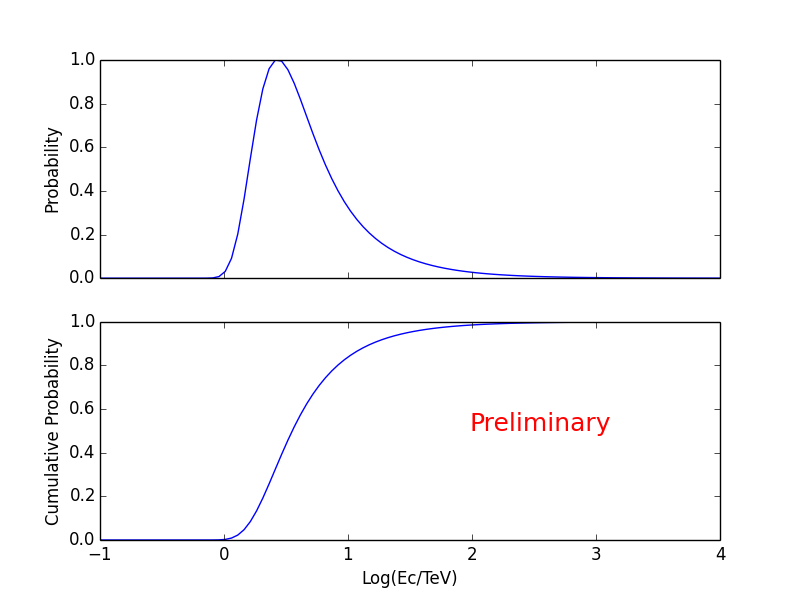}}

  \caption{Left panel: SED of 1ES 0229+200, \FermiLAT-\hess data were fitted
  with 3 different models (PL, LP and PLEC). Right panel: Posterior probability
  and cumulative obtained for 1ES 0229+200 using the bayesian model described in
  this work.}

\label{UL0229}
\end{figure}

Six sources have their \gr\ SED best fitted with a simple power-law model. It is worth
noting that, the found index of the joint \FermiLAT-\hess\ fit is in good
agreement with the results of the \FermiLAT\ analysis (Fig. \ref{UL0229}). This indicates a
continuation of the PL in the intrinsic spectrum of each source and that the
peak position $E_{\rm peak}$ is located at high energy (at or above 100 GeV).

In order to constrain the value of $E_{\rm peak}$, lower limit values have been obtained
using a bayesian approach. For this, the PLEC is used as the spectral model with the collection of data points
(from \hess and \Fermi) $\phi_i$ (measured at an energy $E_i$ with an
uncertainties $\sigma_i$) will be used and noted $Y$.

For this procedure, we define $\Theta$ to be the parameters of our model. The used spectral model being the PLEC then, the evaluated
flux at a given energy is $\Phi(E_i) = f(N,\Gamma,E_{\rm cut})$. This results
that $\Theta = \left\{ N,\Gamma,E_{\rm cut}\right\}$.

Following Bayes' Theorem, it is possible to to write
the posterior probability $P(\Theta|Y)$  as the product of the likelihood
$P(Y|\Theta)$ and the prior probability $P(\Theta)$:

\begin{displaymath}
P(\Theta|Y) \propto P(\Theta)  P(Y|\Theta).
\end{displaymath}

The likelihood $P(Y|\Theta)$ can be written as $$\displaystyle\prod_{i}
P(\phi_i|\Theta) = \displaystyle\prod_{i}
\mathcal{N}(\phi_i|\Phi(E_i),\sigma_i).$$ For the purpose of this model,  each
of the parameters are assumed to be independent, such that the prior $P(\Theta)$
can be expressed as  $$P(\Theta) \propto  \mathcal{N}(N|N_{\rm Fermi},dN_{\rm
Fermi}) \mathcal{N}(\Gamma|\Gamma_{\rm Fermi},d\Gamma_{\rm Fermi}) P(E_{\rm c}).$$
$N_{\rm Fermi}$ and $\Gamma_{\rm Fermi}$ and the associated errors are the
results of the spectral fit of the \FermiLAT\ data with a simple PL. Moreover,
to ensure that this prior tends to zero for large value of $E_{\rm c}$, it this
assumed that $ P(E_{\rm c})\propto 1/E_{\rm c}$.

\begin{figure}
  \centering
  \includegraphics[height=.7\textwidth]{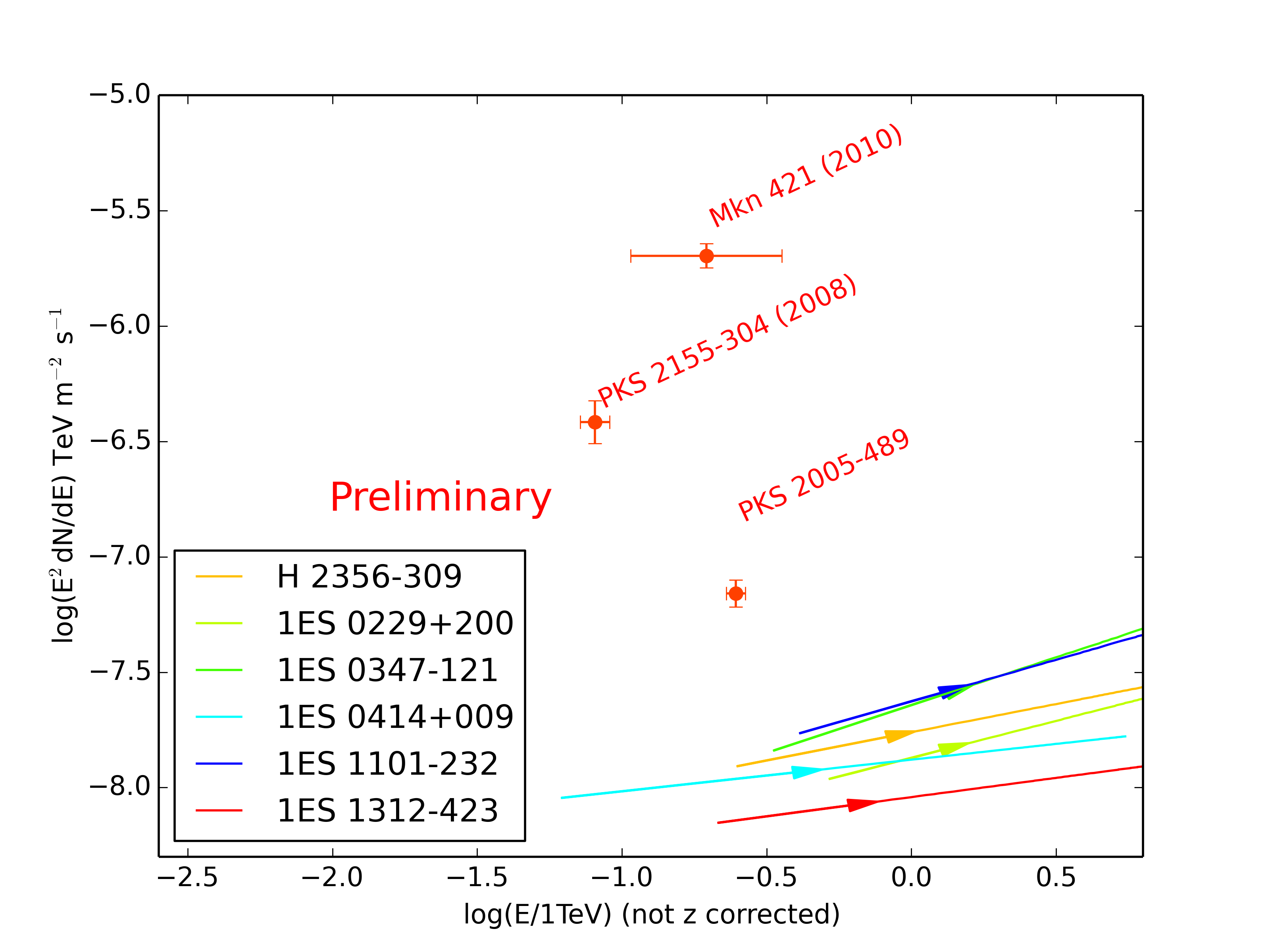}
  \caption{Log of the $E_{\rm peak}$ as a function of the value in the SED
  representation. Sources for which only an lower-limit can be derived, see
  table \ref{table:UL}, an arrow is shown which gives the relation between the
  peak position and the peak value (assuming a PLEC model). }
  \label{Emaxplot}
  \end{figure}

For the computation of the limit, LAT data above 1 GeV and \hess data were used.
Table \ref{table:UL} and Figure \ref{Emaxplot} give the results on position of
the peak at 95 \% confidence level in the $\nu f(\nu)$ representation. The
energy of the peak $E_{\rm peak}$ and value of it ($E^2 dN/dE_{\rm peak}$) are
of course correlated since the lower limit is put on the value of the cut-off
energy. One can note that for most of the sources except  1ES 0414+009, a lower
limit on the peak energy of 200 GeV is obtained with a low flux with respect to
the three sources for which the peak is constrained.

\begin{table*}
\caption{Lower limits at 95\% confidence level obtained on cut-off energy of the PLEC $E_{\rm c}$ with the bayesian model developed in this work. Corresponding peak position $E_{\rm peak}$  and flux in the $\nu f(\nu)$ representation.} 
\label{table:UL} 
\centering %
\begin{tabular}{|l|l|l|l|}\hline
   Name &  $E_{\rm c}$& $E_{\rm peak}$ & $E^2 dN/dE_{\rm peak}$ \\
  &  [TeV]&[TeV]&  $10^{-8}$ [TeV.m$^{-2}$.s$^{-1}$]   \\\hline
   1ES 0229+200 & 1.63 &0.52  & 1.09  \\
   1ES 0347-121 & 0.81 & 0.33& 1.45\\
   1ES 0414+009 & 0.45 & 0.062& 0.90\\
   1ES 1101-232 & 0.45& 0.4& 1.72\\
   1ES 1312-423 & 1.29 & 0.22 & 0.70\\
   H 2356-309   & 1.02 &0.25 & 1.2\\\hline
\end{tabular}
\end{table*}

\section{Summary}

\hess\ and \FermiLAT\ data together with the recent measurement of the EBL density by \hess\ were used to study the intrisic spectra of blazars visible by \hess\ This results in the following findings :

\begin{itemize}
  \item Considering only the \hess\ data, the vast majority of the \hess\ spectra (fitted alone) are well represented by a PL,
  \item Only three sources (the three brightest sources at TeV energies) have their peak in the $\nu f(\nu)$ representation constrained by combined
\Fermi\ and \hess\ data sets,
  \item For the other sources, a limit has been derived on this peak position.
\end{itemize}

\section{Acknowledgements} The support of the Namibian authorities and of the
University of Namibia in facilitating the construction and operation of H.E.S.S.
is gratefully acknowledged, as is the support by the German Ministry for
Education and Research (BMBF), the Max Planck Society, the German Research
Foundation (DFG), the Alexander von Humboldt Foundation, the Deutsche
Forschungsgemeinschaft, the French Ministry for Research, the CNRS-IN2P3 and the
Astroparticle Interdisciplinary Programme of the CNRS, the U.K. Science and
Technology Facilities Council (STFC), the IPNP of the Charles University, the
Czech Science Foundation, the Polish National Science Centre, the South African
Department of Science and Technology and National Research Foundation, the
University of Namibia, the National Commission on Research, Science \& Technology
of Namibia (NCRST), the Innsbruck University, the Austrian Science Fund (FWF),
and the Austrian Federal Ministry for Science, Research and Economy, the
University of Adelaide and the Australian Research Council, the Japan Society
for the Promotion of Science and by the University of Amsterdam. We appreciate
the excellent work of the technical support staff in Berlin, Durham, Hamburg,
Heidelberg, Palaiseau, Paris, Saclay, and in Namibia in the construction and
operation of the equipment. This work benefited from services provided by the
H.E.S.S. Virtual Organisation, supported by the national resource providers of
the EGI Federation.


\begin{thebibliography}{99}
\bibitem{2012A&A...539A.149H} H.E.S.S.~Collaboration, Abramowski, A., Acero, F., et al.\ 2012, A\&A, 539, A149
\bibitem{2013A&A...550A...4H} H.E.S.S.~Collaboration, Abramowski, A., Acero, F., et al.\ 2013, A\&A, 550, A4
\bibitem{2013PhRvD..88j2003A} Abramowski, A., Acero, F., Aharonian, F., et al.\ 2013, PRD, 88, 102003
\bibitem{2016ApJS..223...26A} {Acero}, F., {Ackermann}, M., {Ajello}, M.,{et~al.} 2016, ApJS, 223, 26
\bibitem{3FGL} {Acero}, F., {Ackermann}, M., {Ajello}, M., {et~al.} 2015, ApJS,
218, 23

\bibitem{2007ApJ...664L..71A} Aharonian, F., Akhperjanian, A.~G., Bazer-Bachi,
A.~R., et al.\ 2007, ApJL, 664, L71
\bibitem{2009ApJ...696L.150A} Aharonian, F., Akhperjanian, A.~G., Anton, G., et
al.\ 2009, ApJL, 696, L150
\bibitem{2007NIMPA.583..494A} Albert, J., Aliu, E., Anderhub, H., et al.\ 2007,
Nuclear Instruments and Methods in Physics Research A, 583, 494
\bibitem{2009ApJ...697.1071A} {Atwood}, W.~B., {Abdo}, A.~A., {Ackermann}, M.,
{et~al.} 2009, ApJ, 697, 1071
\bibitem{SummaryObs2} {Dole}, H., {Lagache}, G., {Puget}, J.-L., {et~al.} 2006,
A\&A, 451, 417
\bibitem{2013APh....43..112D} Dwek, E., \& Krennrich, F.\ 2013, Astroparticle Physics, 43, 112
 \bibitem{ebl} {H.E.S.S. collaboration}, Submitted to A\&A
\bibitem{2013arXiv1307.4534S} {Sanchez}, D.~A. \& {Deil}, C. 2013, in
{Proceedings of the 33rd International Cosmic Ray Conference (ICRC 2013)}
\bibitem{2010tsra.confE.197T} Tluczykont, M., \& H.E.S.S.~Collaboration 2010,
25th Texas Symposium on Relativistic Astrophysics, 197


\end{thebibliography}
\end{document}